\documentclass[runningheads,a4paper]{llncs}
\usepackage{subfigure}
\usepackage{multirow}
\usepackage{graphicx}
\usepackage{amssymb}
\usepackage{amsmath}
\usepackage{times}
\usepackage{helvet}
\usepackage{courier}

\usepackage{url}
\urldef{\mailsa}\path|Fangfang.Li@student.uts.edu.au |
\urldef{\mailsb}\path|{Guandong.Xu,Longbing.Cao}@uts.edu.au|
\urldef{\mailsc}\path|{fxz,zniu}@bit.edu.cn|

\begin{document}

\title{Coupled Matrix Factorization within Non-IID Context}
\author{Blind Review}

\maketitle

\begin{abstract}
Recommender systems research has experienced different stages such as from user preference understanding to content analysis. Typical recommendation algorithms were built on the following bases: (1) assuming users and items are IID, namely independent and identically distributed, and (2) focusing on specific aspects such as user preferences or contents. In reality, complex recommendation tasks involve and request (1) personalized outcomes to
tailor heterogeneous subjective preferences; and (2) explicit and implicit objective coupling relationships between users, items, and ratings to be considered as intrinsic forces driving preferences. This inevitably involves the non-IID complexity and the need of combining subjective preference with objective couplings hidden in recommendation applications. In this paper, we propose a novel generic coupled matrix factorization (CMF) model by incorporating non-IID coupling relations between users and items. Such couplings integrate the intra-coupled interactions within an attribute and inter-coupled interactions among different attributes. Experimental results on two open data sets demonstrate that the user/item couplings can be effectively applied in RS and CMF outperforms the benchmark methods.
\end{abstract}

\section{Introduction}
Recommender systems (RS) become increasingly important as they deeply involve our
daily living, online, social, mobile and business activities. Typically, a set of users and items are involved, where each user $u$ rates various items according to his/her respective preferences (embodied by preference rates) \cite{melville_RS}. A new rate or item is then recommended to a user based on the rating behaviors of similar users on existing items.

Often recommendation algorithms come up with the outcomes based on the aggregated
understanding of individual commonality. A rate is then predicted for a new
item to a given user or a new user for a given item. The performance of applying such
algorithms for real-time recommendation for specific users and items is often not very
impressive. There are two important aspect that have not been considered thoroughly in RS.
(1) The heterogeneity between users and between items, namely users and items are personalized and thus rating needs to be tailored according to individual characteristics.
(2) The coupling relationships between users, between items, and between users and
items, namely users and items are coupled and hence rating needs to capture the underlying
interactions. These two aspects together essentially bring the recommendation
problem to a non-IID context, namely users and items are not as independent and
identically distributed (IID) as usually assumed in the existing RS.


The existing RS algorithms and systems such as collaborative filtering and matrix factorization have been mainly built on the IID context, consequently they may overlook or may not fully capture the intrinsic heterogeneity and couplings. For example, many researchers try to influence or precisely estimate the latent factors \cite{Rendle10_factorizationMachine}  \cite{GantnerDFRS10_attributeMapping} \cite{MenonE10_logLinearlatent} \cite{AgarwalC09_regression-mf} \cite{AgarwalC10_flda} \cite{WangB11_collaborative_topic_modeling} through considering the attributes or topics information of users and items for latent factor models. Nevertheless, most of the existing methods assume that the attributes are IID. This is a very fundamental and critical issue for the RS community, as the big recommendation data in online, social, mobile and business applications is essentially non-IID. Specifically, the attributes are more or less interacted and coupled via explicit or implicit relationships \cite{CanWang_IJCAI} \cite{WangSC13} \cite{DBLP:CaoOY12_CBA}. In this paper, we deeply analyse the coupling relationships between users and between items based on their attributes, and incorporate the coupling interactions into MF for filtering the relevant users and items.

\begin{table*}[htbp]
  \centering
  \caption{A Toy Example}
    \begin{tabular}{c c c c |c| c c c c|}
    \cline{5-9}
    &&&&Director & Scorsese & Coppola & Hitchcock & Hitchcock\\
    &&&&Actor & De Niro & De Niro & Stewart & Grant\\
    &&&&Genre & Crime & Crime & Thriller & Thriller\\
    \hline

    \multicolumn{1}{ |c } {Age} & ZipCode & Country & Sex & & God Father & Good Fellas & Vertigo & N by NW\\
    \hline
    \multicolumn{1}{ |c } {20} & 10081 & China & M & $u_1$ & 1 & 3 & 5 & 4 \\
    \multicolumn{1}{ |c } {40} & 2007 & Australia & F & $u_2$ & 4 & 2 & 1 & 5\\
    \multicolumn{1}{ |c } {20} & 2008 & Australia & M & $u_3$ & - & 2 & - & 4\\
    \hline
    \end{tabular}
  \label{tab:toy-example}
\end{table*}

To illustrate the coupling relationships in RS, we give a toy example in  Table \ref{tab:toy-example}. There is a rating matrix consisting of three users and four movies with their attributes. Most existing CF methods utilize the rating matrix for recommendation but ignore the attributes of users and items. However, when the rating matrix is very sparse, the attributes within users and items may also contribute to solving the challenges. Specifically, we can infer the relationship of $u_1$ and $u_2$ from the ``Age", ``ZipCode", ``Country" and ``Sex" attribute space. Similarly, we can get the movies' relationship from the ``Director", ``Actor" and ``Genre" attribute space. Intuitively, the existing similarity methods such as Pearson or Jaccard measures can be applied to compute the similarities within users or items, based on the IID assumption. In reality, however this assumption is not always held and there more or less exist coupling relations between instances and attributes. One observation is that the similarity of two attributes values are dependent on other attributes, for example, two directors' relationship is dependent on ``Actor" and ``Genre" attributes over all the movies. This dependent relation is called the inter-coupled similarity between attributes. Alternatively, within an attribute, one attribute value will also be dependent on other values of the same attribute. Specifically, two attribute values are similar if they present the analogous frequency distribution on one attribute, which leads to another so-called intra-coupled similarity within an attribute. For example, two directors ``Scorsese" and ``Coppola" are considered similar because they appear with the same frequency. We believe that the coupled similarities between values and between attributes should simultaneously contribute to the relationships within users and within items, namely user coupling and item coupling. Incorporating the user coupling and item coupling into the learning MF model may predict more satisfactory recommendations.


The contributions of the paper are concluded as follow:
\begin{itemize}
\item We propose a coupled measure to capture the relationships for users and items, namely user coupling and item coupling, which consider the coupled interaction between attributes from non-IID perspective.
\item We propose a Coupled Matrix Factorization (CMF) model by accommodating the user coupling, item coupling and users' subjective rating preferences together.
\item We conduct experiments to evaluate the superiority of couplings and the effectiveness of CMF model.
\end{itemize}

The rest of the paper is organized as follows. Section 2 presents the related work. In Section 3, we formally state the recommendation and couplings problems. Section 4 first analyses the couplings in RS, then details the coupled MF model integrating the couplings together. Experimental results and analysis are presented in Section 5. The paper is concluded in the last Section.

\section{Related Work}



Collaborative filtering (CF) is one of the most successful approaches taking advantage of user rating history to predict users' interests \cite{SuK09_survey_CF}. As one of the most accurate single models for collaborative filtering, matrix factorization (MF) is a latent factor model \cite{DBLP:Koren08_Factorization} which generally effective at estimating overall structure that relates simultaneously to most items. MF approach tries to decompose the rating matrix to user latent matrix and item latent matrix. Then the estimated rating is predicted by the multiplication of the two decomposed matrices.
With the advent of social network, many researchers have started to analyse social recommender systems and various models integrating social networks have been proposed \cite{DBLP:MaYLK08a_socialRS_PMF}  \cite{DBLP:YangSL12_circleRS}. Social friendship is an outstanding explicit factor to improve the effectiveness of recommendation, however, not every web site have social or trust mechanisms. This explicit social gap strongly motivates us to explore the user and item couplings to improve recommendation qualities. Indeed, such couplings help to make reasonable recommendations when lacking valuable rating information.



Content-based techniques \cite{Pazzani07cbr} are another successful methods which recommend relevant items to users according to users' personal interests. Content-based methods often assume item's attributes are independent which is not always held in reality. Actually, several research outcomes such as \cite{canWang_coupledClustering} \cite{CanWang_IJCAI} \cite{WangSC13} have been proposed to handle the challenging issues. However, to our best of knowledge, limited researches have been done for RS. This motivates us to completely consider the non-IID couplings and integrate the user couplings and item couplings into the MF model.


\section{Problem Statement}

\begin{figure}[!t]
\centering
\includegraphics[scale=0.6]{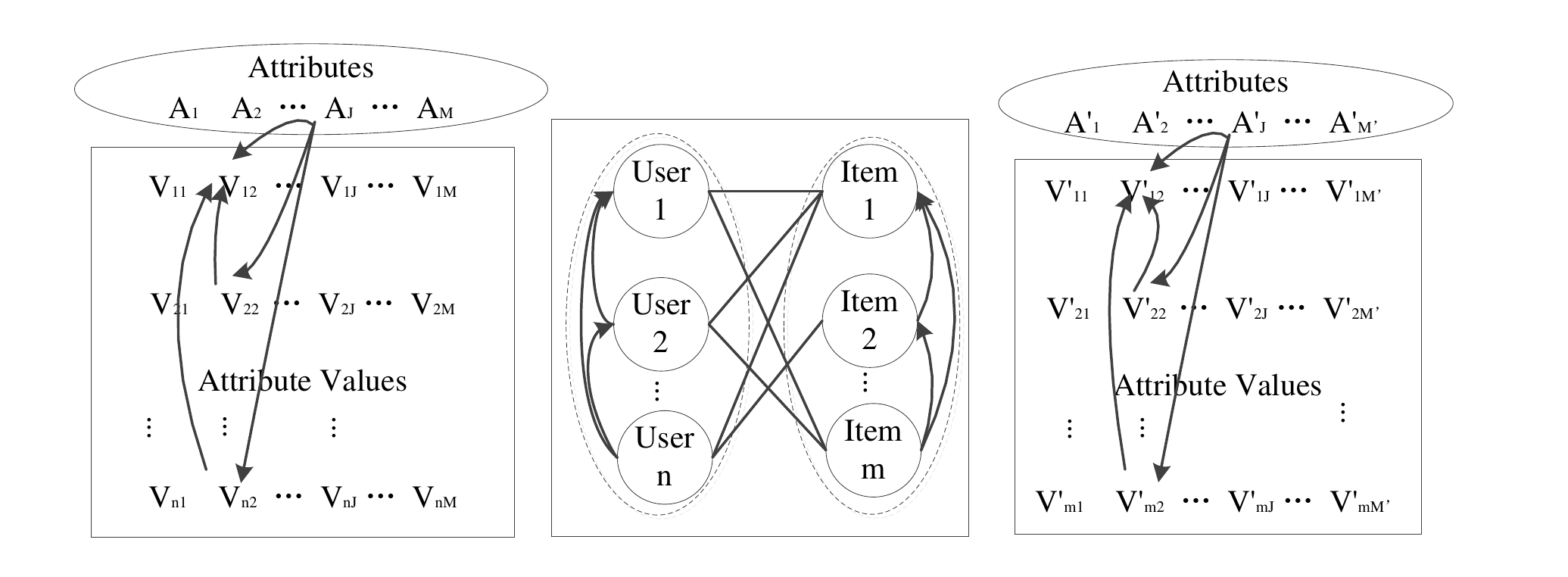}
\caption{Coupling Relations in Recommender Systems}
\label{fig_Coupling}
\end{figure}

A large number of user and item sets with attributes can be organized by a triple $S=<S_U,S_O,h>$, where $S_U = <U,A,V,f>$ describes the users' attribute space, $U=\{u_1, u_2,..., u_n\}$ is a nonempty finite set of users, $A = \{A_1,...,A_M\}$ is a finite set of attributes for users; $V=\cup_{j=1}^J{V_j}{}$ is a set of all attribute values for users, in which $V_j$ is the set of attribute values of attribute $A_j (1\le j \le J)$, $V_{ij}$ is the attribute value of attribute $A_j$ for user $u_i$, and $f=\land_{j=1}^Mf_j (f_j:U \to V_j)$ is an information function which assigns a particular value of each feature to every user. Similar to $S_U$, $S_O = <O,A',V',f'>$ expresses the items' attribute space where $O=\{o_1,...,o_m\}$, $A' = \{A'_1,...,A'_{M'}\}$, $V'=\cup_{j=1}^{J'}{V'_j}{}$, $f'=\land_{j=1}^{M'}{f'}_j ({f'}_j:O \to {V'}_j)$ are all for items. In the triple $S=<S_U,S_O,h>$, $h(u_i,o_j) = r_{ij}$ expresses the subjective rating preference on item $o_j$ for user $u_i$. User rating preferences on items are then converted into a user-item matrix $R$, with $n$ rows and $m$ columns. Each element $r_{ij}$ of $R$ represents the rating given by user $u_i$ on item $o_j$. For instance, Table \ref{tab:toy-example} consists of three users $U=\{u_1,u_2,u_3\}$ and four items $O=\{God Father, Good Fellas, Vertigo, N by NW\}$, $A=\{Age, ZipCode, Country, Sex\}$, $V_3=\{China,Australia\}$, $f_3(u_2)= Australia$, and $A'=\{Director, Actor, Genre\}$, $V'_3=\{Crime,Thriller\}$, $f'_3(Vertigo)= Thriller$, and $h(u_2,Vertigo) = 1$.
The existing similarity methods for computing the relationships assumed that the attributes are independent from each other. However, all the attributes should be coupled together and further influence each other. The couplings are illustrated in Fig.\ref{fig_Coupling}, for users, within an attribute $A_j$, there is dependent relation between values $V_{lj}$ and $V_{mj}$ $(l \not=  m)$. While a value $V_{li}$ of an attribute $A_i$ is further influenced by the values of other attributes $A_j$ $(j \not= i)$. For example, attributes $A_1$, $A_3$, ... to $A_J$ all more or less
influence the values of $V_{12}$ to $V_{n2}$ of attribute $A_2$.

\section{Coupled Matrix Factorization}

\begin{figure*}[!t]
\centerline{\includegraphics[scale=0.6]{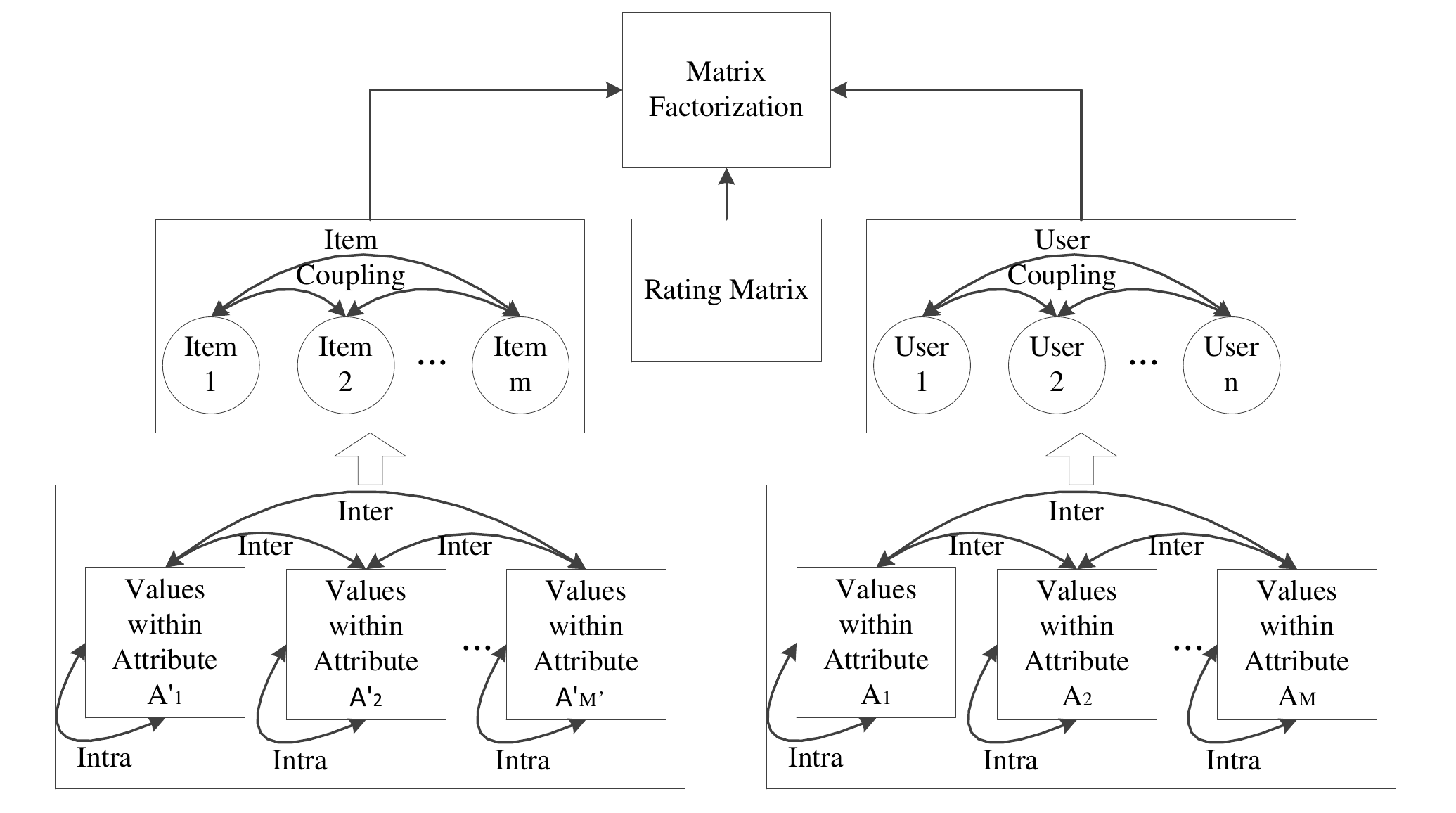}}
\caption{Coupled Matrix Factorization Model}
\label{fig_CoupledFramework-of-MF}
\end{figure*}

In this section, we mainly introduce the coupled MF approach as shown in Fig. \ref{fig_CoupledFramework-of-MF}. CMF first computes the user coupling and item coupling which integrate the coupled interactions based on the objective attributes. Then, user coupling, item coupling and users' rating preferences are incorporated together into MF model.


\subsection{User Coupling}
Users are non-IID, as users share diverse properties but may also be inter-related for some reasons such as educational or cultural background. The user coupling can be calculated on top of the dependent relations for all attributes $A_j$ by setting $S_{Ob}=S_U=<U,A,V,f>$. For two users described by the attribute space, the Coupled User Similarity (\textit{CUS}) is defined by incorporating intra-couplings between values within an attribute and inter-couplings between attributes \cite{canWang_coupledClustering} to measure the similarity between users.

\begin{definition}
Formally, given user attribute space $S_U = <U,A,V,f>$, the \textbf{Coupled User Similarity (\textit{CUS})} between two users $u_i$ and $u_j$ is defined as follows.
\begin{equation}
CUS(u_i,u_j) = \sum\limits_{k = 1}^J {\delta^{Ia}_k(V_{ik}, V_{jk})) * \delta^{Ie}_k(V_{ik}, V_{jk}))}
\end{equation}
where $V_{ik}$ and $V_{jk}$ are the values of attribute $k$ for users $u_i$ and $u_j$, respectively; and $\delta^{Ia}_k$ is the intra-coupling within attribute $A_k$,  $\delta^{Ie}_k$ is the inter-coupling between different attributes.
\end{definition}

\subsection{Item Coupling}
Similarly, items are non-IID. Each item owns different characteristics from others, and there may be coupling relationships between items such as the complementation between purposes. Similar to user coupling, the item coupling can be calculated by setting $S_{Ob}=S_O=<O,A',V',f'>$. For two items described by the attribute space, the Coupled Item Similarity (\textit{CIS}) is defined to measure the similarity between items by integrating intra-couplings within an attribute and inter-couplings between attributes.

\begin{definition}
Formally, given item attribute space $S_O = <O,A',V',f'>$, the \textbf{Coupled Item Similarity (\textit{CIS})} between two items $o_i$ and $o_j$ is defined as follows.
\begin{equation}
CIS(o_i,o_j) = \sum\limits_{k = 1}^{J'} {\delta^{Ia}_k(V'_{ik}, V'_{jk})) * \delta^{Ie}_k(V'_{ik}, V'_{jk}))}
\end{equation}
where $V'_{ik}$ and $V'_{jk}$ are the values of attribute $j$ for items $o_i$ and $o_j$, respectively; and $\delta^{Ia}_k$ is the intra-coupling within attribute $A_k$,  $\delta^{Ie}_k$ is the inter-coupling between different attributes.
\end{definition}

\subsection{Coupled MF Model}
MF approaches have been recognized as the main stream in RS through a latent topic projection learning model. In this work, we attempt to incorporate all discussed couplings into a MF scheme. Traditionally, the matrix of predicted ratings $\hat R\in{\mathbb{R}^{n\times m}}$, where $n$, $m$ respectively denote the number of users and the number of items, can be modeled as: $\hat R = {r_m} + P{Q^T}$
with matrices $P\in{\mathbb{R}^{{n} \times d}}$ and $Q\in{\mathbb{R}^{{m} \times d}}$, where $d$ is the rank (or dimension of the latent space) with $d \le {n},{m}$, and $r_m \in \mathbb{R}$ is a global offset value. Through this modelling, the prediction task of matrix $\hat{R}$ is transferred to compute the mapping of items and users to factor matrices $P$ and $Q$. Once this mapping is completed, $\hat R$ can be easily reconstructed to predict the rating given by one user to an item.

In our proposed CMF, we take not only the rating matrix, but also the user coupling and item coupling, into account. All these aspects should be accommodated into a unified learning model. The learning procedure is constrained by three-fold: the learned rating values should be as close as possible to the observed rating values, the predicted user and item profiles should be similar to their neighbourhoods as well, which are derived from their coupling information. Specifically, in order to incorporate the user coupling and item coupling, we add two additional regularization factors in the optimization step. Then the computation of the mapping can be similarly optimized by minimizing the regularized squared error. The objective function is given as Eqn. \ref{objctiveFuction-CMF}.
\begin{equation}\label{objctiveFuction-CMF}
\begin{split}
& L = \frac{1}{2}\mathop \sum \limits_{\left( {u,o_i} \right) \in K} {\left( {{R_{u,o_i}} - \hat R_{u,o_i}} \right)^2} +  \frac{\lambda}{2}\left( {\|{Q_i\|^2} + \|{P_u\|^2}} \right)+ \frac{\alpha }{2} \sum\limits_{all(u)} \\& {{{\left\| {{P_u} - \sum\limits_{v \in {\Bbb N}(u)} {{CUS(u,v)}{P_v}} } \right\|}^2}}  + \frac{\beta }{2}\sum\limits_{all(o_i)} {{{\left\| {{Q_i} - \sum\limits_{o_j \in {\Bbb N}(o_i)} {{CIS(o_i,o_j)}{Q_j}} } \right\|}^2}}
\end{split}
\end{equation}

As we can see in the objective function, the rating preference, user coupling and item coupling have been all incorporated together. Specifically, the first part reflects the subjective rating preferences and the latter two parts reflect the user coupling and item coupling, respectively. This means when we recommend relevant items to users, the users' rating preferences may take the dominant role. Besides this, another distinct advantage is that, when we do not have ample rating data, it is still possible to make satisfactory recommendations via leveraging the coupling information, e.g., one user will be recommended what his/her neighbours like or items similar to what he/she preferred before.

To optimize the above objective equation, we minimize the objective function $L$ by the gradient descent approach:
\begin{equation}\label{gradientDescentP}
\begin{split}
& \frac{{\partial L}}{{\partial {P_u}}} = \sum\limits_{o_i} {{I_{u,o_i}}} ({r_m} + {P_u}{Q_i}^T - {R_{u,o_i}}){Q_i} + \lambda {P_u} +
 \alpha ({P_u} - \\& \sum\limits_{v \in {\Bbb N}(u)} {{CUS(u,v)}{P_v}} ) -  \alpha \sum\limits_{v:u \in {\Bbb N}(v)} {{CUS(u,v)}({P_v} - \sum\limits_{w \in {\Bbb N}(v)} {{CUS(v,w)}{P_w}} )}
\end{split}
\end{equation}

\begin{equation}\label{gradientDescentQ}
\begin{split}
& \frac{{\partial L}}{{\partial {Q_i}}} = \sum\limits_u {{I_{u,o_i}}} ({r_m} + {P_u}{Q_i}^T - {R_{u,o_i}}){P_u} + \lambda {Q_i}+ \beta ({Q_i} - \sum\limits_{o_j \in {\Bbb N}(o_i)} \\&{{CIS(o_i,o_j)}{Q_j}} ) - \beta \sum\limits_{o_j:o_i \in {\Bbb N}(o_j)} {{CIS(o_j,o_i)}({Q_j} - \sum\limits_{o_k \in {\Bbb N}(o_j)} {{CIS(o_j,o_k)}{Q_k}} )}
\end{split}
\end{equation}
where $I_{u,o_i}$ is the function indicating that whether user has rated item $o_i$, 1 means rated, 0 means not rated. $CUS(u,v)$ is the coupled similarity of users $u$ and $v$, and $CIS(o_i,o_j)$ is the coupled similarity of items $o_i$ and $o_j$. ${\Bbb N}(u)$ and ${\Bbb N}(o_i)$ respectively represent the user and item neighborhood filtered by coupled similarity.

\subsubsection{Model Training}
Through the above gradient descent approach, the best matrices $P$ and $Q$ can be computed in terms of the user coupling, item coupling and user-item coupling. The whole process of the coupled model starts at computing user coupling and item coupling based on their objective attribute space, then neighbors of users and items are selected from the couplings. Next, values of $P$ and $Q$ are randomly initiated followed by an iteration step to update $P$ and $Q$ until convergence according to Eqn. \ref{gradientDescentP} and \ref{gradientDescentQ}. After $P$ and $Q$ are learned from the training process, we can predict the ratings for user-item pairs $(u,o_i)$.

\section{Experiments and Results}
In this section, we evaluate our proposed model and compare it with the existing approaches respectively using Movielens\footnote{www.movielens.org} and Bookcrossing\footnote{www.bookcrossing.com} data sets.
\subsection{Data Sets}
Movielens data set has been widely explored in RS research in last decade. Movielens 1M data set consists of 1,000,209 anonymous ratings of approximately 3,900 movies made by 6,040 Movielens users who joined Movielens in 2000. Only users providing basic demographic information such as ``gender", ``age", ``occupation" and ``zipcode" are included in this data set. The movies also have a special ``genre" attribute which is applied for computing the item couplings.

Similarly, collected by Cai-Nicolas Ziegler, Bookcrossing data set contains 278,858 users with demographic information providing 1,149,780 ratings on 271,379 books. The ratings range from 1 to 10 and the users' ``gender" and ``age" attributes and the books' ``book-author", ``year of publication" and ``publisher" have been used to form user and item couplings.

\subsection{Experimental Settings}
The 5-fold cross validation is performed in our experiments. In each fold, we have 80\% of data as the training set and the remaining 20\% as testing set. Here we use Root Mean Square Error (RMSE) and Mean Absolute Error (MAE) as evaluation metrics.

%
%

To evaluate the performance of our proposed CMF we consider five baseline approaches:
(1) The basic probabilistic matrix factorization (PMF) approach \cite{SalMnih08_PMF};
(2) Singular value decomposition (RSVD) \cite{svd-gene} is a factorization method to decompose the rating matrix;
(3) Implicit social matrix factorization (ISMF) \cite{DBLP:implicit_social_SIGIR13} is an unified model which incorporates implicit social relationships between users and between items computed by Pearson similarity based on the user-item rating matrix;
(4)	User-based CF (UBCF) \cite{SuK09_survey_CF} first computes users' similarity by Pearson Correlation on the rating matrix, then recommends relevant items to the given user according to the users who have strong relationships;
(5)	Item-based CF (IBCF) \cite{Deshpande:2004:ITN_recommendation} first considers items' similarity by Pearson Correlation on the rating matrix, then recommends relevant items which have strong relationships with the given user's interested items.

The above five baselines just consider users' rating preferences on items but ignore the attributes of users and items. In order to demonstrate the effectiveness of our proposed method, we also compare it with other three hybrid models PSMF, CSMF and JSMF which respectively augment MF with Pearson Correlation Coefficient, Cosine and Jaccard similarity measures to compute the relationships within users and items based on their attributes. Simply put, we first respectively apply Pearson Correlation Coefficient, Cosine and Jaccard similarity to compute the similarities for users and items based on their attributes. Then we utilize the similarities within users and items to update and optimize the objective function to acquire the best $P$ and $Q$ as CMF. The objective fuction of PSMF, CSMF and JSMF is given as Enq. \ref{objctiveFuction-PSMF}. $Sim(u,v)$ respectively represents the pearson similarity, cosine similarity, and jaccard similarity between users $u$ and $v$, and $Sim(o_i,o_j)$ represents the pearson, consine and jaccard similarities between items $o_i$ and $o_j$.

\begin{equation}\label{objctiveFuction-PSMF}
\begin{split}
& L = \frac{1}{2}\mathop \sum \limits_{\left( {u,o_i} \right) \in K} {\left( {{R_{u,o_i}} - \hat R_{u,o_i}} \right)^2} +  \frac{\lambda}{2}\left( {\|{Q_i\|^2} + \|{P_u\|^2}} \right)+ \frac{\alpha }{2}\sum\limits_{all(u)} \\& {{{\left\| {{P_u} - \sum\limits_{v \in {\Bbb N}(u)} {{Sim(u,v)}{P_v}} } \right\|}^2}}  +\frac{\beta }{2}\sum\limits_{all(o_i)} {{{\left\| {{Q_i} - \sum\limits_{o_j \in {\Bbb N}(o_i)} {{Sim(o_i,o_j)}{Q_j}} } \right\|}^2}}
\end{split}
\end{equation}

%
%

\subsection{Experimental Results and Discussions}
We respectively evaluate the effectiveness of our CMF model in comparison with the above baselines and the other three hybrid methods PSMF, CSMF and JSMF by selecting the optimal parameters $\alpha$=1.0, $\beta$=0.2 for Movielens, and $\alpha$=0.6, $\beta$=1.0 for Bookcrossing.

\subsubsection{Superiority over MF Methods}
Because the users of Movielens data set have the basic demographic data for user coupling, and the movies also have natural genre attribute for item coupling, the experimental results compared to the MF methods in Table \ref{tab:comparisons_mf} can show the effect of user coupling and item coupling. In this experiment, we respectively compared the experimental results regarding different latent dimensions with MAE and RMSE metrics. When the latent dimension is set as 100, 50 and 10, in terms of MAE, our proposed CMF can reach an average improvements of 21.24\%, 12.88\% compared with PMF and RSVD approaches. Similarly, CMF can also improves averagely 58.77\%, 39.93\% regarding RMSE over PMF and RSVD approaches. Besides the basic comparisons, we also compare our CMF with the latest research outcome ISMF which utilizes the implicit relationships between users and items based on the rating matrix by Pearson similarity. From the experimental result, we can see that CMF can averagely improve 15.22\% and 45.87\% regarding MAE and RMSE respectively. In conclusion, the experiments on Movielens data set clearly indicate that CMF is more effective than the baseline MF approaches and the state-of-the-art ISMF method regarding MAE and RMSE when latent dimension is respectively set to 100, 50 and 10, due to the strength of user coupling and item coupling.
\begin{table*}[hbp]
  \centering
  \caption{MF Comparisons on Movielens and Bookcrossing}
    \begin{tabular}{|c|c|c|c|c|c||c|}
    \hline
    Data Set & Dim & Metrics & PMF (Improve) & ISMF (Improve)  & RSVD (Improve)  & CMF \\ \hline
    \multirow{6}{*}{Movielens} & \multirow{2}{*}{100D} & MAE & 1.1787(28.09\%) & 1.1125 (21.47\%) & 1.1076 (20.98\%) & \textbf{0.8978}\\ \cline{3-7}

    & & RMSE &1.7111 (71.07\%) & 1.5918 (59.14\%) & 1.5834 (58.30\%) & \textbf{1.0004}\\\cline{2-7}

    & \multirow{2}{*}{50D} & MAE &1.1852 (18.43\%) & 1.1188 (11.79\%) & 1.1088 (10.79\%) & \textbf{1.0009}\\\cline{3-7}

    & & RMSE &1.8051 (58.98\%) & 1.6103 (39.50\%) & 1.5835 (36.82\%) & \textbf{1.2153}\\ \cline{2-7}
	
    & \multirow{2}{*}{10D} & MAE &1.2129 (17.19\%) & 1.1651 (12.41\%) & 1.1098 (6.88\%) & \textbf{1.0410}\\ \cline{3-7}

    & & RMSE &1.8022 (46.25\%) & 1.7294 (38.97\%) & 1.5863 (24.66\%)& \textbf{1.3397}\\ \cline{2-7}
    \hline
     \hline

    \multirow{6}{*}{Bookcrossing} & \multirow{2}{*}{100D} & MAE &1.5127 (3.65\%) & 1.5102 (3.40\%) & 1.5131 (3.69\%)  & \textbf{1.4762}\\ \cline{3-7}
    & & RMSE &3.7455 (0.76\%) & 3.7397 (0.18\%)& 3.7646 (2.67\%) & \textbf{3.7379}\\ \cline{2-7}

    & \multirow{2}{*}{50D} & MAE &1.5128 (3.67\%) & 1.5100 (3.39\%) & 1.5131 (3.70\%) & \textbf{1.4761}\\\cline{3-7}
    & & RMSE &3.7452 (0.74\%) & 3.7415 (0.37\%) & 3.7648 (2.70\%) & \textbf{3.7378}\\ \cline{2-7}

    & \multirow{2}{*}{10D} & MAE &1.5135 (3.73\%) & 1.5107 (3.45\%) & 1.5134 (3.72\%) &  \textbf{1.4762}\\ \cline{3-7}
    & & RMSE &3.7483 (1.20\%) & 3.7440 (0.77\%) & 3.7659 (2.96\%) & \textbf{3.7363}\\\cline{2-7}
    \hline
    \end{tabular}
  \label{tab:comparisons_mf}
\end{table*}

Similar to Movielens, Bookcrossing data set also has certain user demographic information, and rich book content information such as ``Book-Title", ``Book-Author", ``Year-Of-Publication" and ``Publisher". After removing all the invalid ISBNs, all the books in the data set are cleaned. Therefore the experimental results on the Bookcrossing data set can also demonstrate the impacts of user and item couplings. We depict the effectiveness comparisons with respect to different methods on Bookcrossing data set in Table \ref{tab:comparisons_mf}. We can clearly see that, our proposed CMF method outperforms all the counterparts in terms of MAE and RMSE. Specifically, when the latent dimension is set as 100, 50 and 10, in terms of MAE, our proposed CMF can reach an average improvements of 3.68\%, 3.70\% compared with PMF and RSVD approaches. While CMF can averagely increase 0.98\% and 2.78\% regarding RMSE over PMF and RSVD approach. Furthermore, the prominent improvements compared with the baseline approaches regarding MAE and RMSE are resulted from considering complete couplings. Additionally, we also compare the CMF with the state-of-the-art method ISMF, the result shows that the improvements can reach to 3.41\% and 0.44\% regarding MAE and RMSE respectively. Therefore, we can conclude that our CMF method not only outperforms PMF and SVD which are basic MF methods, but also performs better than the state-of-the-art model ISMF in terms of MAE and RMSE metrics.

\subsubsection{Superiority over CF Methods}

In addition to the MF methods, we also compare our proposed CMF model with two different CF methods UBCF and IBCF. In this experiment, we fix the latent dimension to 100 for our proposed CMF model. On Movielens, the results in Table \ref{tab:comparisons_cfs} indicate that CMF can respectively improve 0.49\% and 2.42\% regarding MAE, and 0.18\% and 19.54\% in terms of RMSE. Similarly compared with UBCF and IBCF, on Bookcrossing data set, the results show that the CMF can reach huge improvements respectively 33.02\% and 31.03\% regarding MAE, and 24.68\% and 19.04\% regarding RMSE. Therefore, this experiment clearly demonstrates that our proposed CMF performs better than UBCF and IBCF methods. The improvements are contributed by the full consideration of the couplings in RS.
\begin{table*}[hbp]
\centering
  \caption{CF Comparisons on Movielens and Bookcrossing}
  \label{tab:comparisons_cfs}
\begin{tabular}{|c|c|c|c||c|}\hline
Data Set & Metrics & UBCF (Improve)  & IBCF (Improve) & CMF\\\hline
\multirow{2}{*}{Movielens} & MAE & 0.9027 (0.49\%)& 0.9220 (2.42\%) & \textbf{0.8978}\\\cline{2-5}
& RMSE & 1.0022 (0.18\%) & 1.1958 (19.54\%) & \textbf{1.0004}\\
\hline
\hline
\multirow{2}{*}{Bookcrossing} & MAE & 1.8064 (33.02\%) & 1.7865	(31.03\%) &\textbf{1.4762}\\\cline{2-5}
& RMSE & 3.9847	(24.68\%) &	3.9283 (19.04\%) & \textbf{3.7379}\\\hline
\end{tabular}
\end{table*}

\subsubsection{Superiority over Hybrid Methods}
\begin{figure}
  \centering
  \subfigure[MAE on Movielens]{
    \label{fig:subfig-strategy-mae-movielens} 
    \includegraphics[width=0.42\textwidth]{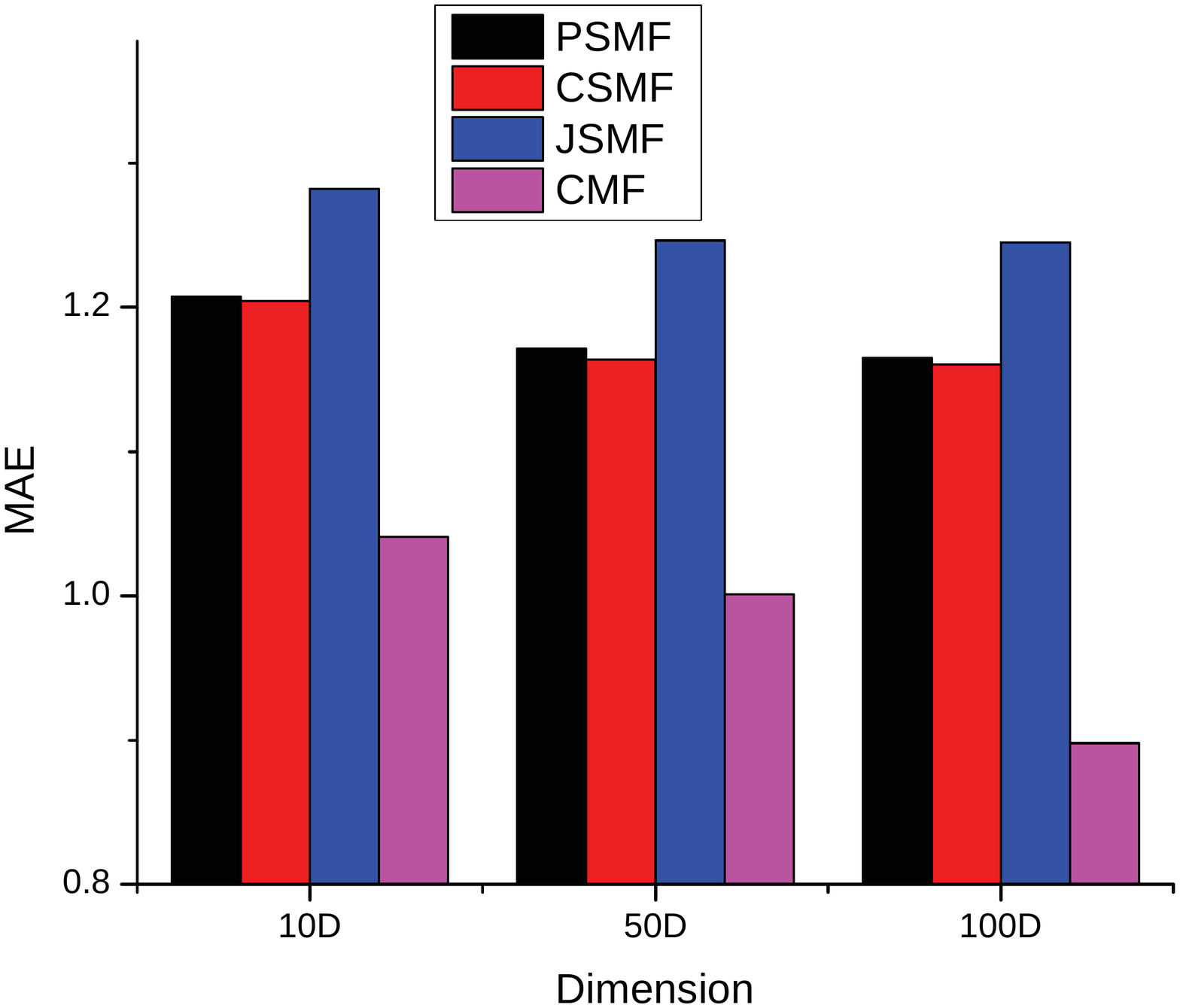}}
  \subfigure[RMSE on Movielens]{
    \label{fig:subfig-strategy-rmse-movielens} 
    \includegraphics[width=0.42\textwidth]{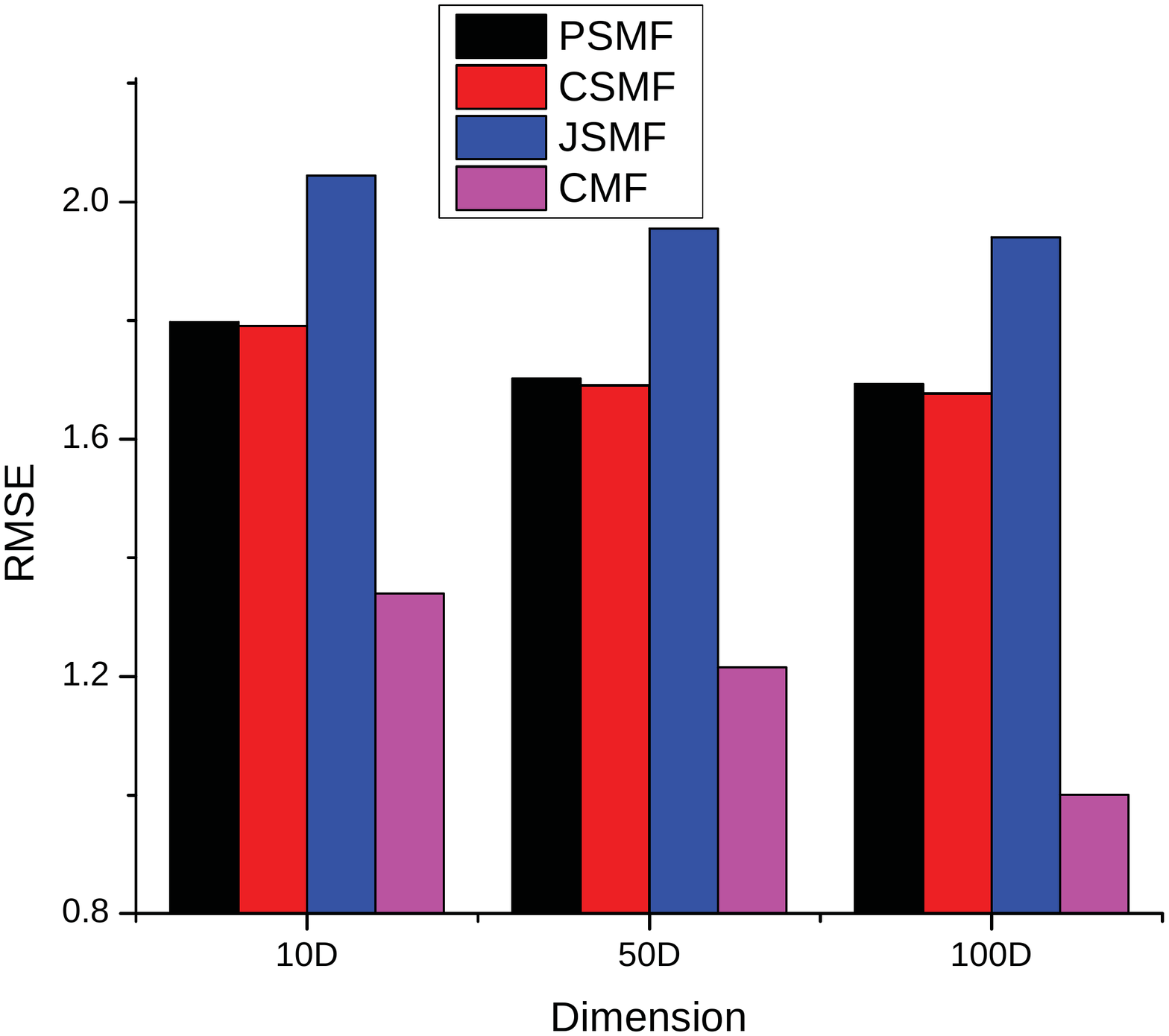}}
      \subfigure[MAE on Bookcrossing]{
    \label{fig:subfig-strategy-mae} 
    \includegraphics[width=0.42\textwidth]{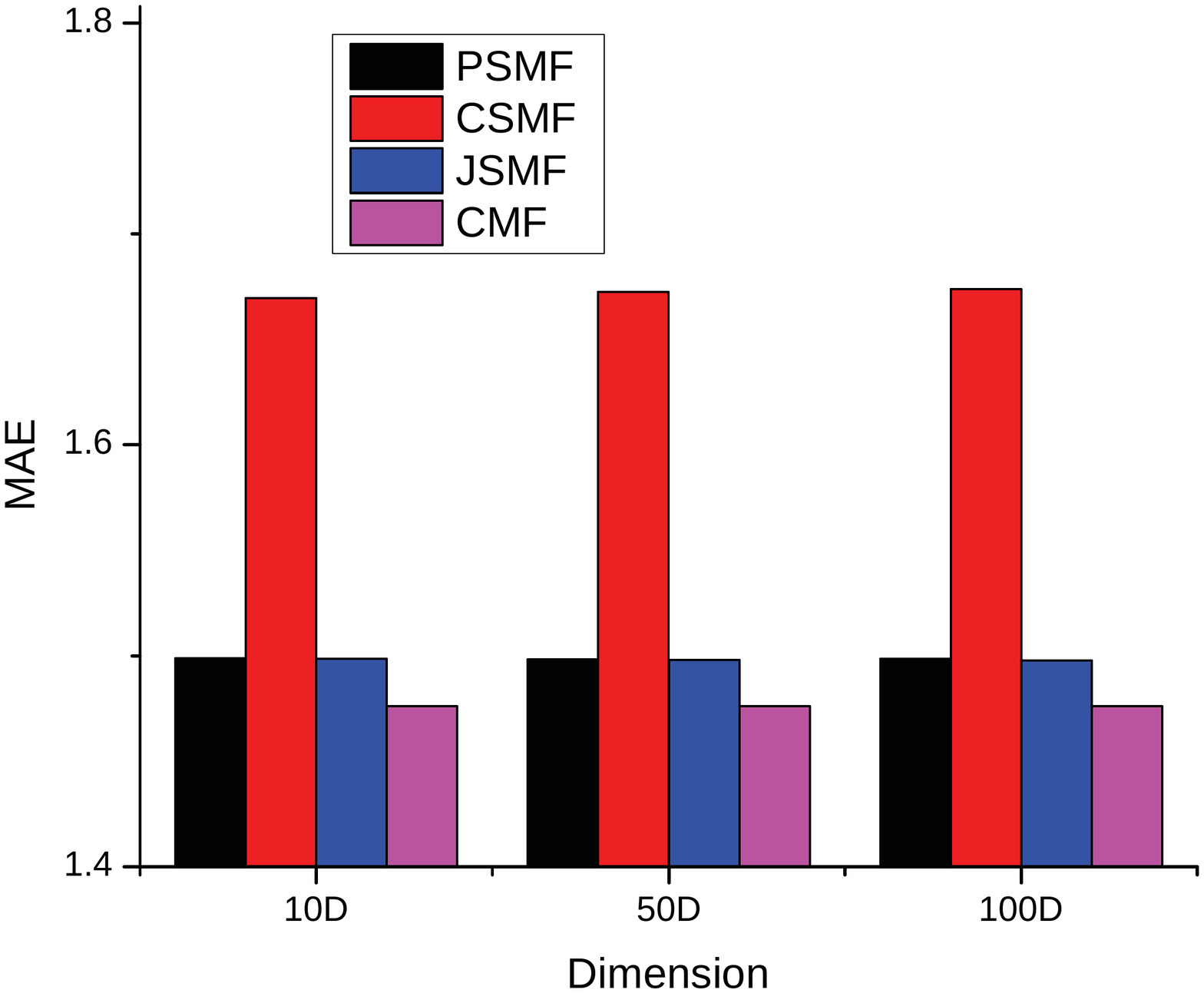}}
  \subfigure[RMSE on Bookcrossing]{
    \label{fig:subfig-strategy-rmse} 
    \includegraphics[width=0.42\textwidth]{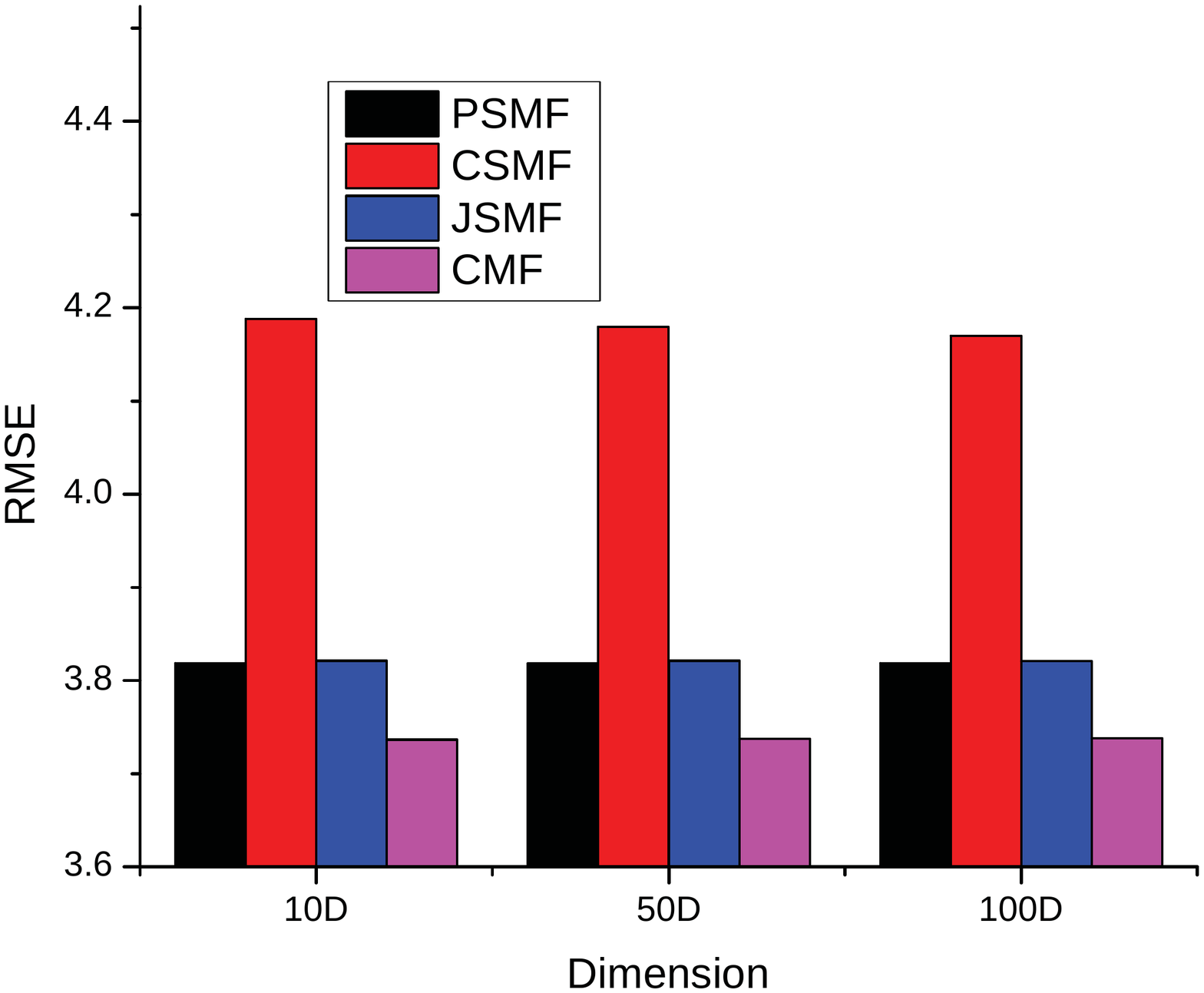}}
  \caption{Superiority over Hybrid Methods on Movielens and Bookcrossing}
  \label{fig-strategy-ml-bookcrossing} 
\end{figure}

In order to demonstrate the effectiveness of our proposed method, we also compare it with other three hybrid methods  PSMF, CSMF and JSMF. From the resultant Fig. \ref{fig-strategy-ml-bookcrossing} on Movielens data set, we can clearly see that the proposed CMF method greatly outperforms PSMF, CSMF and JSMF in terms of MAE and RMSE. Specifically, CMF can averagely improve PSMF, CSMF, JSMF by 20.14\%, 19.63\%, 27.78\% regarding MAE, and by 54.58\%, 53.45\%, 79.50\% regarding RMSE. Similarly on Bookcrossing data set, the results in Fig. \ref{fig-strategy-ml-bookcrossing} clearly indicate that the CMF method also perform better than PSMF, CSMF and JSMF regarding MAE and RMSE. The results show that CMF can respectively improve 2.24\%, 19.57\%, 2.2\% in average regarding MAE, and 8.13\%, 44.18\%, 8.38\% in terms of RSME compared with PSMF, CSMF, JSMF. From this experiment, we can conclude that our proposed CMF is more effective than these three hybrid methods.

\section{Conclusion and Future Work}
In this paper, we studied Recommender System from a non-IID perspective, specifically, we mainly focused on the significant non-IID coupling relations between users and between items to improve the quality of recommendations. The couplings disclosed the traditional IID assumption and deeply analysed the intrinsic relationships between users and between items. Furthermore, a coupled matrix factorization model was proposed to incorporate the coupling relations and the explicit rating information. The experiments conducted on the real data sets demonstrated the superiority of the proposed CMF method and suggested that non-IID couplings can be effectively applied in RS. The heterogeneity between users and between items is still not thoroughly considered in this paper. We need to further explore the heterogeneity challenge for enhancing our recommendation model in the future.

%


\bibliography{ref}
\bibliographystyle{plain}

\end{document}